\documentclass[%
 reprint,
superscriptaddress,
 amsmath,amssymb,
 aps,
]{revtex4-1}

\usepackage{subcaption}
\usepackage{amsmath}
\usepackage{graphicx}
\usepackage{dcolumn}
\usepackage{bm}
\usepackage{hyperref}

\begin{document}

\preprint{APS/123-QED}

\title{Testing Einstein Maxwell Power-Yang-Mills Hair via Black Hole Photon Rings
}%

\author{Zuting Luo}
\author{Meirong Tang}
 \author{Zhaoyi Xu}
 \email{Electronic address:zyxu@gzu.edu.cn (Corresponding author)
}
\affiliation{%
College of Physics, Guizhou University,\\
 Guiyang 550025, China
}%





\date{\today}

\begin{abstract}In this paper, the optical appearance of static and spherically symmetric hairy black holes is studied under the standard Einstein-Maxwell theory considering the p-power Yang-Mills term. During the research process, the specific case of $p=1/2$ was mainly selected for discussion. To understand the impact of the hairy parameter on black holes, we have studied the event horizon radius $r_{h} $, the photon sphere radius $r_{ph}$  and the radius of the innermost stable circular orbit $r_{isco}$  of this hairy black hole. Then, we utilized the backward ray-tracing method to analyze the geodesics of photons around this black hole and discussed the influence of the hairy parameter on the photon geodesics. In addition, we also calculated the unique shadow and photon ring of the black hole irradiated by a static thin accretion disk with three toy model emission functions. The research results show that as the hairy parameter gradually increases, the event horizon radius  $r_{h} $, the photon sphere radius  $r_{ph}$, the radius of the innermost stable circular orbit  $r_{isco}$ and the critical impact parameter $b_{ph}$ of the black hole all exhibit a decreasing trend. Meanwhile, it also causes the area of the black hole shadow and the photon ring to decrease accordingly. Consequently, in the case of the static and spherically symmetric standard Einstein Maxwell power-Yang-Mills hairy black hole, there is no degeneracy in the photon ring and the shadow. Theoretically, it can reflect different black hole solutions and thus verify the Yang-Mills hair.
\begin{description}
\item[Keywords]
Black hole, Shadows, Photon rings, Null geodesics,Yang-Mills hair.
\end{description}
\end{abstract}

\maketitle

\section{Introduction}Although general relativity (GR) has achieved great success in classical scenarios, it still faces significant challenges under extreme conditions \cite{Penrose:1964wq,Hawking:1970zqf}. The singularity theorems of Penrose and Hawking indicate that GR loses its predictive ability in certain regions, which prompts physicists to propose many alternative theories of gravity or consider quantum effects to eliminate singularities.\cite{Rovelli:2004tv,Brunnemann:2005in,Modesto:2007wp,Bodendorfer:2011nx,Long:2019nkf,Long:2020wuj,Bojowald:2001xe}. The Einstein-Cartan theory avoids the appearance of singularities by introducing torsion\cite{Hehl:1976kj}; quantum gravity frameworks such as string theory and loop quantum gravity provide a completely new explanatory perspective for avoiding the problem of spacetime singularities \cite{Green:1982tc,Han:2005km,Ashtekar:2011ni,Zhang:2022vsl,Gambini:2008dy,Ashtekar:2011ni}. In addition, the Einstein-Yang-Mills theory, by unifying non-Abelian gauge fields and gravity within the geometrized framework of curved spacetime, provides a new tool for studying the interactions between gauge fields and gravitational fields in complex backgrounds\cite{Jackiw:1976pf,Brink:1976bc,Dorigoni:2024vrb,Koutrolikos:2024neu}. This theory not only expands the scope of application of the GR but also has great significance in describing asymptotic symmetries, black hole physics and the research on quantum gravity\cite{Barnich:2013sxa}. The proposal of alternative theories of gravity has provided new ideas for GR to overcome its limitations under extreme conditions.

black hole, as one of the classic predictions of GR, is not only an important tool for verifying existing gravitational theories but also a crucial experimental object for testing the validity of alternative theories and distinguishing differences among theories. In recent years, the exploration of black holes has always been at the forefront of theoretical physics research. With the release of the images of the supermassive black hole M87* by the Event Horizon Telescope (EHT) in 2019 \cite{EventHorizonTelescope:2019dse,EventHorizonTelescope:2019uob,EventHorizonTelescope:2019jan,EventHorizonTelescope:2019ths,EventHorizonTelescope:2019pgp,EventHorizonTelescope:2019ggy} and the images of the supermassive black hole SgrA* at the center of the Milky Way in 2022 \cite{EventHorizonTelescope:2022wkp}, the existence of black holes, the mysterious celestial bodies predicted by the GR, has also been proven. It can be clearly seen from the pictures released by the EHT that there is an obviously dark region in the central part of the black hole. Usually, this specific dark region is named the shadow of the black hole. The shadow is the region inside the critical curve, which is closely related to the radially unstable spherical photon orbit \cite{Guo:2020qwk}. Meanwhile, the shadow is also a prominent feature of the image of the accretion flow around supermassive black holes \cite{Johannsen:2013vgc}. The shape and size of the shadow depend on the spacetime metric. The research on black hole shadows provides us with an important method to extract effective information about black hole spacetime and helps us test various alternative theories of gravity. After Synge's pioneering work \cite{Synge:1966okc}, a large number of studies on black hole shadows have been carried out in various gravitational theories \cite{Johannsen:2010ru,Li:2021riw,Sui:2023tje,Abdujabbarov:2016hnw,Wei:2013kza,Mizuno:2018lxz,Konoplya:2019sns,Zare:2024dtf}.

There is a bright ring surrounding the black hole shadow. This bright ring is called the photon ring\cite{Wang:2023vcv}. It is an important concept in the research on black hole imaging. Through the strong gravitational lensing effect, the light around the black hole is repeatedly deflected around it, and finally forms photon trajectories one after another \cite{Wang:2022yvi}. These photon trajectories together present a "ring-shaped" image around the black hole. This phenomenon not only provides crucial clues for black hole imaging but also serves as an important tool for studying physical phenomena in the extreme environment near black holes. Through observations of the photon ring, people have gained a deeper understanding of the properties of black holes such as the structure of the accretion disk, the gravitational lensing effect, and the rotation speed of black holes \cite{Sui:2023tje,Gan:2021xdl,Meng:2024puu,Yang:2022btw}. The photon ring provides a method for testing the no-hair theorem of black holes \cite{Yang:2024utv} and studying the gravitational effects of black holes \cite{Gralla:2019xty}, as well as a way to extract information about black hole parameters such as mass, spin, and event horizon size \cite{Gogoi:2024vcx}. In addition, people's research on the higher-order ring images of black hole accretion disks has also made it possible to use them to distinguish different accretion models \cite{Bisnovatyi-Kogan:2022ujt}. Therefore, the observational images of the photon rings and shadows of black holes are an effective tool \cite{Rincon:2023hvd}, which can help us better understand the properties of black holes. 

The no-hair theorem indicates that usually a black hole can be described by three physical parameters, namely mass, angular momentum, and electric charge. This implies that although the internal conditions before the formation of a black hole may be extremely complicated, it becomes a relatively simple celestial body after its formation. However, for modified theories of gravity, the introduction of modification terms adds extra degrees of freedom, which can explain phenomena that are difficult to be fully explained by the GR. In fact, in the process of exploring modified theories of gravity, people are committed to introducing additional matter fields into the theoretical system. The introduction of these additional matter fields has brought about a significant result, that is, a black hole can no longer be characterized solely by the traditional three parameters but requires more "hairy" parameters to fully describe its properties. It is precisely based on such a background that hairy black holes, as a new research object, have been widely constructed and deeply analyzed in related fields \cite{Sui:2023tje,Hawking:2016msc,Herdeiro:2014goa,Antoniou:2017acq,Antoniou:2017hxj,Hod:2011aa}.

Recently, Gabriel Gómez et al. obtained the spherically symmetric solution of the Einstein Maxwell theory in the presence of the Yang-Mills field with a power-term\cite{Gomez:2023qyv}. This spherically symmetric black hole solution includes the interaction between the Yang-Mills field and the mass-charge of the black hole. We hope to use the black hole photon ring to explore how the Yang-Mills field really has an impact on the black hole spacetime. Specifically, we focus on the model with power terms $p=1/2$ and $Q=0.6$. This special case represents a black hole possessing both Abelian  charges $Q$ and non-Abelian charges  $q_{YM}$  and is asymptotically non-flat. When our Yang-Mills hair parameter $Q_{YM}$  is set to zero, it reverts to the standard Reissner-Nordström (RN) black hole solution. Therefore, this paper aims to carry out research by analyzing this model and leveraging the structural characteristics of the photon ring and shadow mentioned earlier. Since there are differences in the structural characteristics of the photon ring and shadow among different black hole solutions, it is expected that the changes in these characteristics can be used to distinguish different black hole solutions, thereby testing the Yang-Mills hair.  

The organizational structure of this paper is as follows: In Section \ref{dierjie}, the Einstein Maxwell power-Yang-Mills black hole solution is presented. Meanwhile, the value situation of the event horizon of this black hole under the action of the hairy parameter $Q_{YM}$ is explored. In Section \ref{disanjie}, the geodesic equation of the hairy black hole is first derived. Then the effective potential and the photon sphere radius are discussed. In Section \ref{disijie}, the light bending around the black hole is analyzed. Assuming the existence of an optically and geometrically thin accretion disk around the black hole, the geodesic trajectories under the influence of different hairy parameters are given. Moreover, the total emission and total observed intensities as well as the optical appearance under three different toy models with different values of the hairy parameter \(Q_{YM}\) are studied and compared. Section \ref{diwujie} is the summary. In addition, throughout this paper, the natural unit system is adopted, i.e., \(c = G = 1\), and \(M = 1\) is set. The metric signature \((-, +, +, +)\) is used in this paper. 

\section{the Einstein Maxwell Power-Yang-Mills Black Hole}\label{dierjie}
In this section, first briefly introduce the static and spherically symmetric hairy black hole under the standard Einstein-Maxwell theory with the p-power Yang-Mills term. We consider the spherically symmetric spacetime (in Schwarzschild coordinates), and the line element of this black hole can be written as\cite{Gomez:2023qyv}
\begin{equation}\label{1}
	\begin{split}
		d s^{2}=-f(r) d t^{2}+f(r)^{-1} d r^{2}+r^{2}\left(d \theta^{2}+\sin ^{2} \theta d \varphi^{2}\right)
	\end{split},
\end{equation}
where
\begin{equation}\label{2}
	\begin{split}
		f(r)=1-\frac{2 M}{r}+\frac{Q^{2}}{r^{2}}+\frac{Q_{Y M}}{r^{4 p-1}}
	\end{split}.
\end{equation}
This solution can be understood as a linear combination of the Yang-Mills term and the standard RN solution,and the Yang-Mills charge  $q_{YM}$ is related to its normalized version as\cite{Mazharimousavi:2009mb}
\begin{equation}\label{3}
	\begin{split}
		Q_{YM} \equiv \frac{2p - 1}{4p - 3} q_{YM}^{2p}
	\end{split}.
\end{equation}
It can be seen from Eq. \eqref{2} that, compared with the standardized RN black hole solution, the hole metric in this paper only introduces a nonlinear parameter $p$  and a hairy parameter $Q_{YM}$. Their introduction indicates the geometric deformation on the radial and time metric components. In this paper, the cases of $p=1/2$ and the Abelian charge $Q=0.6$ are taken. Therefore, when the hairy parameter $Q_{YM}=0$, it can revert to the standard RN black hole solution. Since this metric is static and spherically symmetric, and the event horizon, as a hypersurface with spacetime symmetry, should also be static and spherically symmetric. Therefore, when solving for the event horizon, it should only be a function of $r$, independent of $t$, $\theta$ and $\varphi$. That is to say, the event horizon should satisfy
\begin{equation}\label{4}
	\begin{split}
g^{rr} = 0
	\end{split},
\end{equation}
we have
\begin{equation}\label{5}
	\begin{split}
		f(r) = 1 - \frac{2M}{r} + \frac{Q^2}{r^2} + \frac{Q_{YM}}{r^{4p - 1}} = 0
	\end{split}.
\end{equation}
When we determine the values of the power term $p=1/2$ and the Abelian charge $Q=0.6$, solving Eq. (5) can yield the horizon situations under different $Q_{YM}$, and they are shown in Fig. \ref{1}. The black dashed line in the figure represents the position of the event horizon radius of the RN black hole. We find that when $Q_{YM}=1.77778$, there is only one root corresponding to the Einstein Maxwell power-Yang-Mills black hole having only one horizon. When $Q_{YM}$is less than this value, the black hole will have two horizons; conversely, when $Q_{YM}$ is greater than this value, there is no horizon.
\begin{figure}[h]
	\includegraphics[width=0.5\textwidth]{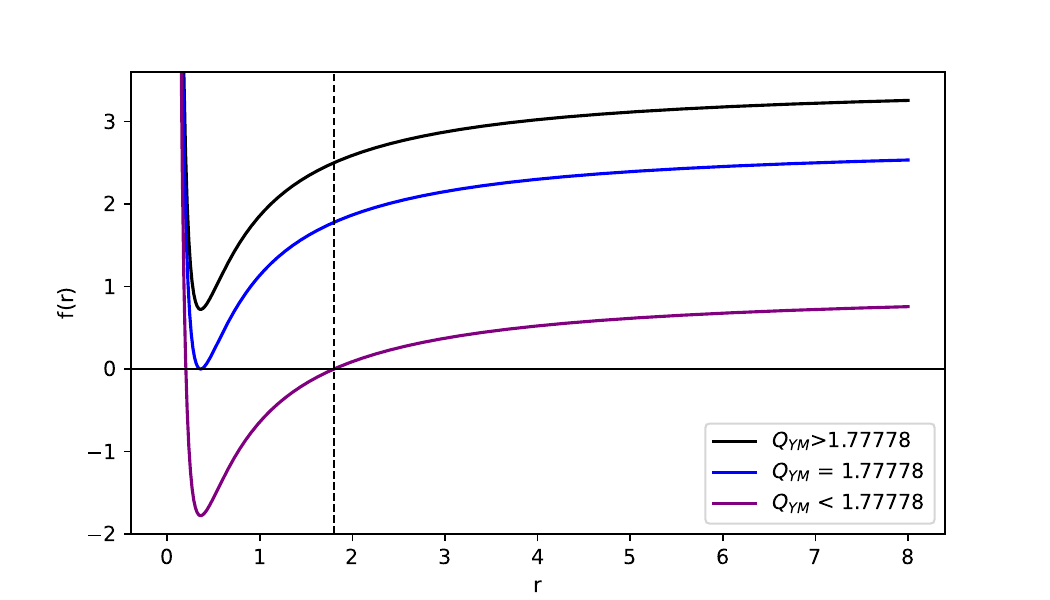}
	\caption{The situation of the metric function for different values of $Q_{YM}$.} 
	\label{1}
\end{figure}
\section{Null Geodesics in the Hairy Black Hole Spacetime}\label{disanjie}
In order to study the influence of the hairy parameter $Q_{YM}$ of the Einstein Maxwell power-Yang-Mills black hole on the photon ring and the shadow, we need to first study the equation of motion of photons in the spacetime of this black hole and discuss the effective potential of photons. In the spherically symmetric spacetime, we consider the photon motion constrained to a fixed plane, $\theta = \pi/2$, that is
\begin{equation}\label{6}
	\begin{split}
		\sin \theta = 1, \frac{d \theta}{d \tau} = 0
	\end{split}.
\end{equation}
For the spacetime of the line element \eqref{1}, the geodesic equation can be derived according to the Lagrangian  
\begin{equation}\label{7}
	\begin{split}
	2 \mathcal{L} = g_{ij} \frac{dx^i}{d\tau} \frac{dx^j}{d\tau}	
	\end{split},
\end{equation}
 where $\tau$ is the affine parameter along the geodesic. From the spacetime metric, the Lagrangian is
\begin{equation}\label{8}
	\begin{split}
		\mathcal{L} = \frac{1}{2} \left[ - f(r) \dot{t}^2 + \frac{1}{f(r)} \dot{r}^2 + r^2 \dot{\theta}^2+ (r^2 \sin^2 \theta) \dot{\varphi}^2 \right]
	\end{split}.
\end{equation}
Where the dot indicates the derivative with respect to $\tau$, and the corresponding canonical momenta are
\begin{equation}\label{9}
	\begin{split}
			p_{t} &= \frac{\partial \mathcal{L}}{\partial \dot{t}} = - f(r) \dot{t},\\
		p_{r} &= - \frac{\partial \mathcal{L}}{\partial \dot{r}} = - \frac{1}{f(r)} \dot{r},\\
		p_{\theta} &= - \frac{\partial \mathcal{L}}{\partial \dot{\theta}} = - r^{2} \dot{\theta},\\
		p_{\varphi} &= - \frac{\partial \mathcal{L}}{\partial \dot{\varphi}} = - r^{2} \sin^{2} \theta \dot{\varphi}.	
	\end{split}
\end{equation}
In the Schwarzschild spacetime, there are two Killing vector fields, namely the timelike Killing vector field $\left( \frac{\partial}{\partial t} \right)^a$ and the axial Killing vector field $\left( \frac{\partial}{\partial \varphi} \right)^a$. These two Killing vector fields will lead to two conserved quantities, that is, the conservation of energy \(E\) and the conservation of angular momentum \(L\)\cite{Gralla:2019xty}.
\begin{equation}\label{10}
	\begin{split}
	E = - g_{00} \frac{d t}{d \tau} = f(r) \frac{d t}{d \tau}	
	\end{split},
\end{equation}
\begin{equation}\label{11}
	\begin{split}
		L = g_{33} \frac{d \varphi}{d \tau} = r^2 \frac{d \varphi}{d \tau}
	\end{split}.
\end{equation}
Substituting  Eq. \eqref{10} and \eqref{11} into \eqref{8}, we get:
\begin{equation}\label{12}
	\begin{split}
	2 \mathcal{L} = -\frac{E^2}{f(r)} + \frac{1}{f(r)} \dot{r}^2 + \frac{L^2}{r^2}
	\end{split}.
\end{equation}
Also, because the Lagrangian of the null geodesic is zero:
\begin{equation}\label{13}
	\begin{split}
		\frac{1}{f(r)} \left( \frac{d r}{d \tau} \right)^2 = \frac{E^2}{f(r)} - \frac{L^2}{r^2}
	\end{split}.
\end{equation}
From Eq. \eqref{13}, an ordinary differential equation for the radius $r$ with respect to the azimuthal angle $\varphi$ in the orbital plane can be further obtained\cite{Gogoi:2024vcx}
\begin{equation}\label{14}
	\begin{split}
		\left( \frac{d r}{d \varphi} \right)^2 = \frac{1}{b^2} r^4 - f(r) r^2 \equiv V(r)
	\end{split}.
\end{equation}
Here, $b = L/E$ is the impact parameter. It can be seen that the photon trajectory is determined by the impact parameter $b$. In order to study the hole shadow, we need to determine the radius of the unstable photon sphere, which is determined by the following equation\cite{Yang:2022btw,Gogoi:2023ntt}
\begin{equation}\label{15}
	\begin{split}
		\left. V(r) \right|_{r = r_{ph}} = 0
	\end{split},
\end{equation}
\begin{equation}\label{16}
	\begin{split}
		\left. \frac{dV(r)}{dr} \right|_{r = r_{ph}} = 0.
	\end{split}
\end{equation}
Due to spherical symmetry, Eq. \eqref{16} can be simplified to
\begin{equation}\label{17}
	\begin{split}
		\left. \frac{d}{dr} \frac{f(r)}{r^2} \right|_{r = r_{ph}} = 0
	\end{split}.
\end{equation}
After obtaining the radius of the photon sphere from Eq. \eqref{17}, substituting it into Eq. \eqref{15} will yield the corresponding critical impact parameter as
\begin{equation}\label{18}
	\begin{split}
		b_{ph} = \frac{r_{ph}}{\sqrt{f(r_{ph})}}
	\end{split}.
\end{equation}
The effective potential for the radial motion of photons can be used to distinguish different photon orbits, and the effective potential at the radial distance $r$ can be expressed as\cite{Gogoi:2024vcx}
\begin{equation}\label{19}
	\begin{split}
		V_{\text{eff}}(r) \equiv f(r) \frac{1}{r^2}
	\end{split}.
\end{equation}
Therefore, the effective potential at the photon sphere is
\begin{equation}\label{20}
	\begin{split}
	V_{\text{eff}}(r_{ph}) = \frac{1}{b_{ph}^2}
	\end{split}.
\end{equation}
Above, $r_{ph}$ is the radius of the photon sphere, and $b_{ph}$ is the corresponding critical impact parameter. Their relationship is shown in Fig. \ref{2}. Here, the case corresponding to a RN black hole is taken as an example.
\begin{figure}[h]
	\includegraphics[width=0.5\textwidth]{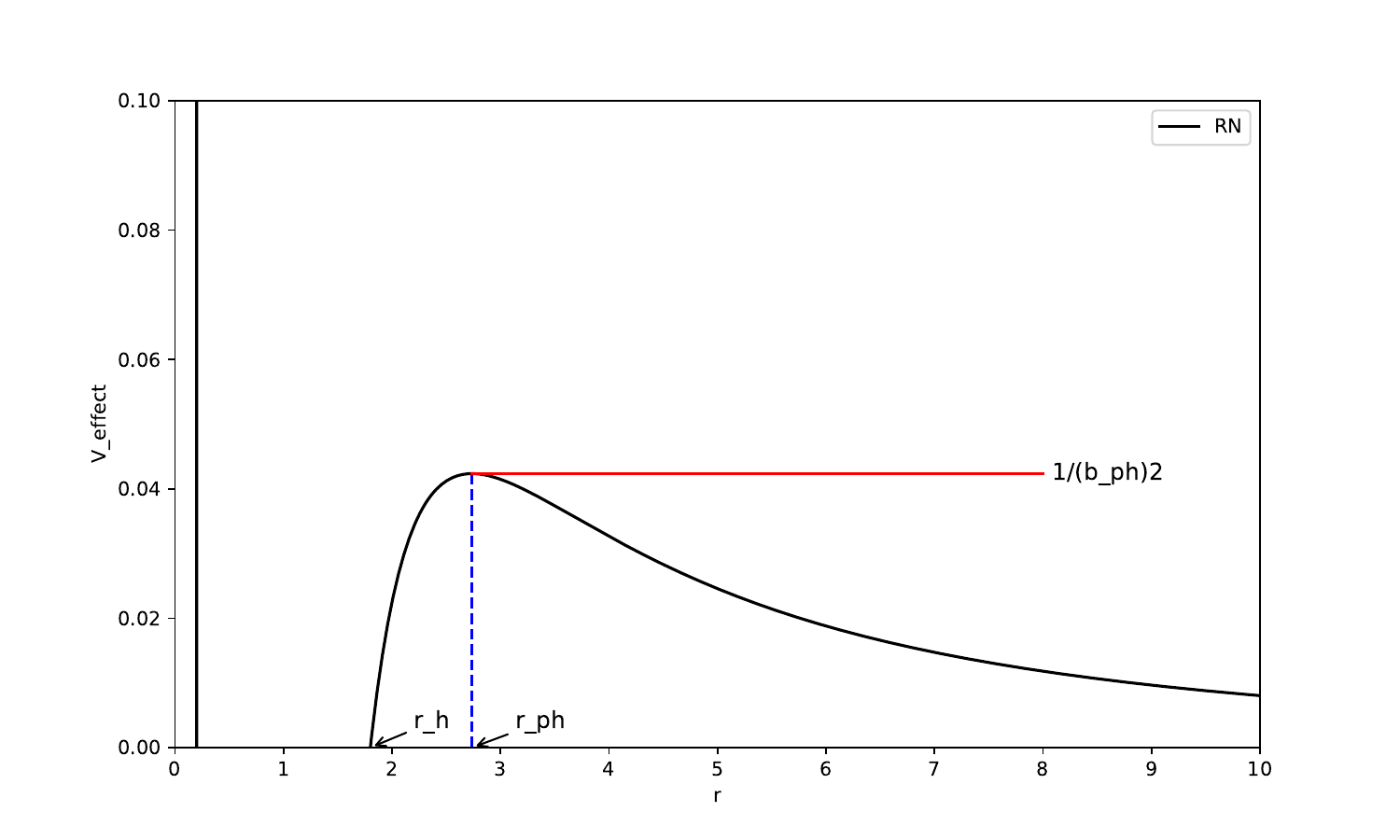}
	\caption{This is the curve of the variation of the effective potential $V_{\text{eff}}(r)$ of an RN black hole with the charge $Q=0.6$ as a function of the radius $r$, as well as the relationship between the photon sphere radius $r_{ph}$ and the critical impact parameter $b_{ph}$.}
	\label{2}
\end{figure}
In the figure, for an RN black hole with $Q=0.6$, the event horizon radius is $r_{h}=1.8$, the photon sphere radius is $r_{ph}=2.73693$ and the critical impact parameter is $b_{ph}=1.85869$. The maximum value in the figure represents the radius of the unstable photon sphere. Further, we show the photon effective potential images for different hairy parameters $Q_{YM}$ in Fig. \ref{3} . It is not difficult to see from Fig. \ref{3} that as  $Q_{YM}$ increases, the photon effective potential will gradually increase, and the event horizon radius of the black hole and the photon sphere radius of the photons will also decrease.
\begin{figure}[h]
	\includegraphics[width=0.5\textwidth]{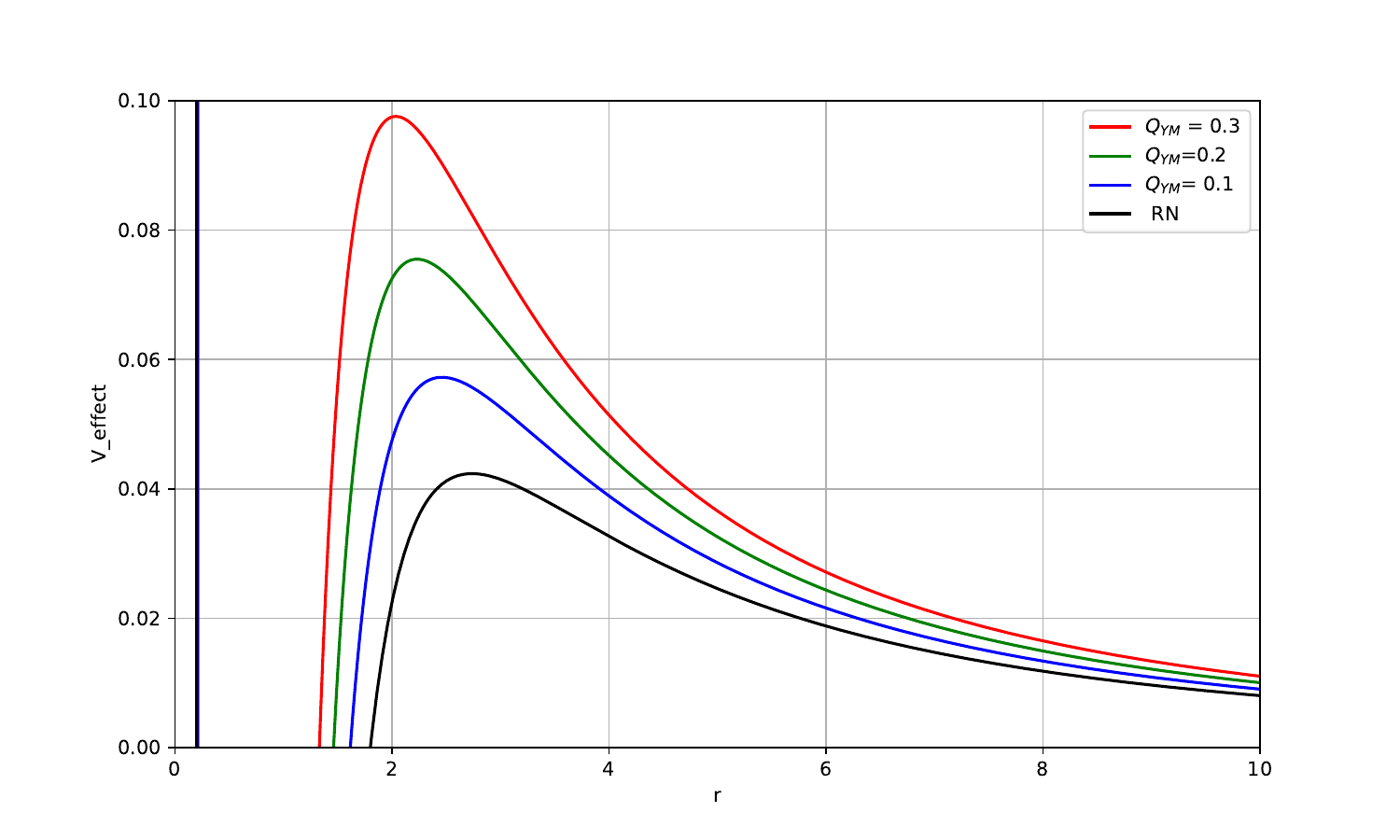}
	\caption{This is the curve of the variation of the effective potential $V_{\text{eff}}(r)$ with $r$ when the electric charge $Q=0.6$ and the hairy parameter $Q_{YM}$ takes different values, with the cases of $Q_{YM}=0.1$, $Q_{YM}=0.2$, $Q_{YM}=0.3$ and RN taken respectively.}
	\label{3}
\end{figure}

The value of the effective potential is of crucial significance in the study of photon orbits. When the effective potential reaches its maximum value, the corresponding orbit is an unstable photon sphere. Once photons on such an orbit are slightly disturbed, they may either fall into the black hole or escape to infinity and be captured by distant observers. Conversely, when the effective potential takes on a minimum value, the orbit at this time is stable, allowing photons to move continuously and stably on the corresponding orbit. And the unstable photon sphere that can be transmitted to distant observers has an important influence on observing the characteristics of the accretion image\cite{Gan:2021pwu}. Therefore, in the following discussion, we will focus on the unstable photon sphere orbits.
\section{The photon ring and shadow of the Einstein Maxwell power-Yang-Mills black hole}\label{disijie}
The fine structure of the photon ring involves the complex behaviors of photon orbits near black holes, especially the distribution of photons on different orbits. Since photons on different orbits appear as layers with different brightnesses when imaged, it helps observers identify the characteristics of black holes. In this part, we will adopt the method of backward ray tracing to analyze the behaviors of null geodesics near the Einstein Maxwell power-Yang-Mills hairy black hole and further explore how the optical appearance of the black hole presents when a very thin accretion disk serves as a light source. Here, the accretion disk is idealized as an optically and geometrically extremely thin plane, that is, the disk surface has no absorption or scattering effects on photons. During the research process, we assume that the background radiation has no impact on the results and that the accretion disk is the only light source. To simplify the analysis, we fix the accretion disk on the equatorial plane and assume that the observer is located in the direction of the North Pole at the same time. Due to the strong gravitational pull of the black hole, the null geodesics of photons may intersect with the accretion disk multiple times before falling into the black hole or escaping to infinity. The difference in the number of such intersections will have an important impact on the total light intensity received by the observer. Therefore, we need to classify these light rays according to the number of intersections between the null geodesics and the accretion disk and calculate the imaging effect of the Einstein Maxwell power-Yang-Mills hairy black hole based on this. 
\subsection{ Null Geodesics: Direct emission, lensing rings, and photon rings}
To facilitate the study of the motion trajectories of photons near black holes, we perform a transformation on Eq. \eqref{14} with $r = \frac{1}{u}$, so that the geodesic equation of photons becomes: 
\begin{equation}\label{21}
	\begin{split}
		\left( \frac{du}{d\varphi} \right)^2 = \frac{1}{b^2}-f\left( \frac{1}{u} \right)u^2
	\end{split}.
\end{equation}
It can be known from the geodesic Eq. \eqref{21} that the motion trajectory of photons is closely related to the impact parameter $b$, as described in the literature \cite{Gralla:2019xty}. For all null geodesics that are parallelly directed towards the North Pole, the range of their impact parameters depends on the number of intersections between the null geodesics and the accretion disk, and the number of intersections can be expressed as:
\begin{equation}\label{22}
	\begin{split}
		n = \frac{\varphi}{2 \pi}
	\end{split}.
\end{equation}
In the formula, $\varphi$ is the total change in the polar angle in the polar coordinate system during the completion of the entire trajectory of the null geodesic. From the photon trajectory equation of Eq. \eqref{21}, when the condition of $b < b_{\text{ph}}$ is met, the total change in the polar angle outside the event horizon can be expressed by Eq. \eqref{23}\cite{Yang:2022btw}
\begin{equation}\label{23}
	\begin{split}
		\varphi = \int_0^{u_h} \frac{1}{\sqrt{\frac{1}{b^2}-f\left( \frac{1}{u} \right)u^2}} du
	\end{split}.
\end{equation}
In the formula, $u_h \equiv 1/r_h$ and $r_{h}$ is the radius of the event horizon. When $b > b_{\text{ph}}$, the total change in the polar angle is
\begin{equation}\label{24}
	\begin{split}
		\varphi = 2 \int_0^{u_{\text{max}}} \frac{1}{\sqrt{\frac{1}{b^2}-f\left( \frac{1}{u} \right)u^2}} du
	\end{split}.
\end{equation}
Among them, $u_{\text{max}} \equiv 1/r_{\text{min}}$, and $r_{min}$ is the minimum radial distance from the position of the light ray on its trajectory to the black hole. In order to discuss the observational images of radiation near a black hole, it is necessary to clarify that the observed intensity is closely related to the number of intersections between geodesics and the accretion disk, and the number of intersections $n$ is a function of $b$ satisfying such a relationship\cite{Peng:2020wun}.
\begin{equation}\label{25}
	\begin{split}
	n(b)=\frac{2m - 1}{4}, \quad m = 1,2,3,\cdots.
	\end{split}
\end{equation}
It can be known from Eq. \eqref{25} that, for a given $m$, we use $b_m^{\pm}$ to represent two different solutions\cite{Gralla:2019xty,Peng:2020wun}, where $b_m^{-} < b_c$ and $b_m^{+} > b_c$. Then, we can classify all trajectories as follows:

(1) Direct emission:\(\frac{1}{4}<n<\frac{3}{4}\Rightarrow b\in(0,b_{2}^{-})\cup(b_{2}^{+},\infty)\);

(2) Lens ring:\(\frac{3}{4}<n<\frac{5}{4}\Rightarrow b\in(b_{2}^{-},b_{3}^{-})\cup(b_{3}^{+},b_{2}^{+})\);

(3) Photon ring:\(n > \frac{5}{4}\Rightarrow b\in(b_{3}^{-},b_{3}^{+})\).
\begin{figure}[h]
	\includegraphics[width=0.5\textwidth]{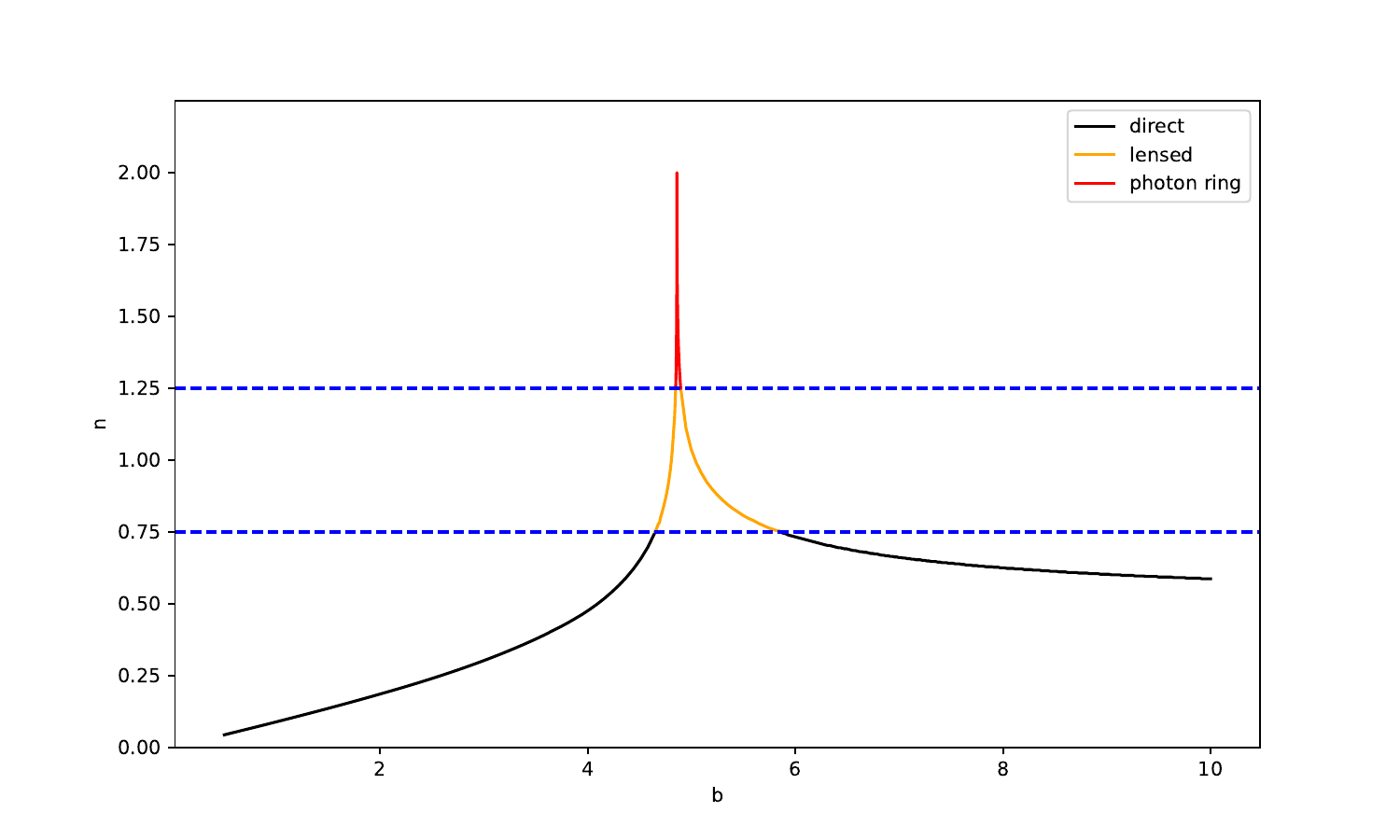}
	\caption{The variation curve of the number of intersections $n$ with the impact parameter $b$ for an RN black hole when $Q=0.6$. The black line represents direct emission, the yellow line represents the lensing ring, and the red line represents the photon ring.}
	\label{4}
\end{figure}

We present in Fig. \ref{4} the variation curve of the number of intersections $n$ of null geodesics of an RN black hole with the accretion disk with respect to the impact parameter $b$ when $Q=0.6$. The black line in the figure represents the photon trajectory of direct emission, the yellow line represents the lensing ring, and the red line represents the photon ring. It is not difficult to see that when the impact parameter is less than the critical impact parameter, the number of intersections increases as the impact parameter increases, reaching a maximum at the critical parameter, and the opposite is true when the impact parameter is greater than the critical impact parameter. Similarly, using the null geodesic equation of photons Eq. \eqref{21}, we present in Fig. \ref{5} the variation relationship between the number of intersections $n$ of null geodesics with the accretion disk and the impact parameter $b$ for different values of the hairy parameter $Q_{YM}$ when $Q=0.6$. It can be seen from Fig. \ref{5} that the variation trend of the number of intersections $n$ of null geodesics with the accretion disk with respect to the impact parameter $b$ for different values of $Q_{YM}$ is similar to that of the RN case. However, as $Q_{YM}$ increases, the entire variation curve shifts to the left, which corresponds to the fact that the critical impact parameter gradually decreases as $Q_{YM}$ increases. Moreover, the impact parameter intervals corresponding to the lensing ring and the photon ring also gradually decrease as $Q_{YM}$ increases.
\begin{figure}[h]
	\includegraphics[width=0.5\textwidth]{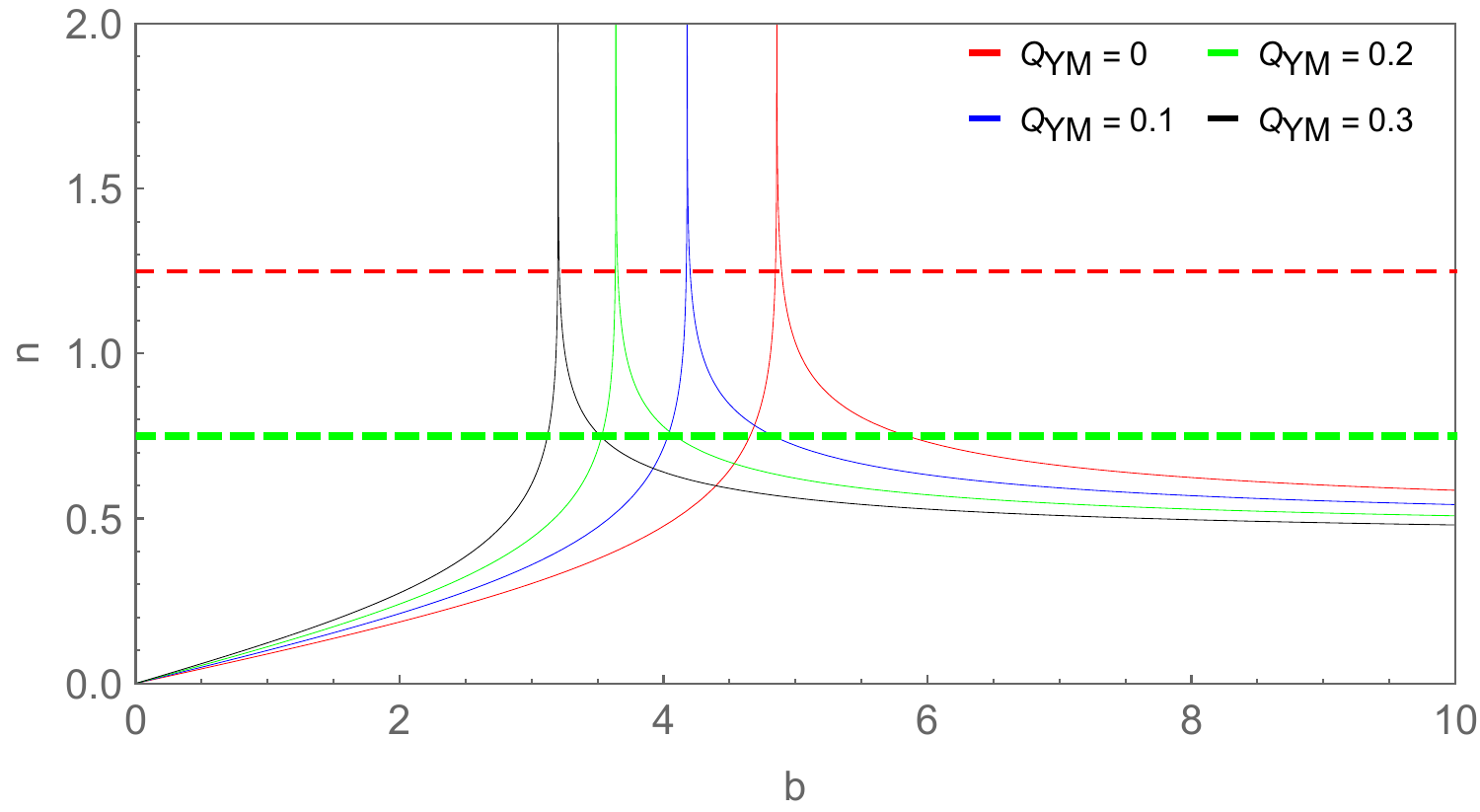}
	\caption{The curve of the variation relationship between the number of intersections $n$ of null geodesics with the accretion disk and the impact parameter $b$ for different values of the hairy parameter $Q_{YM}$. In the figure, four cases are taken with $Q_{YM}$ being 0.3 (black), 0.2 (green), 0.1 (blue), and 0 (red).}
	\label{5}
\end{figure}

In addition, by using the null geodesic Eq. \eqref{21}, we have drawn the light ray trajectory diagrams near the black hole for different values of $Q_{YM}$ in Fig. \ref{6}.
\begin{figure}[h]
	\centering
	\includegraphics[width=0.22\textwidth]{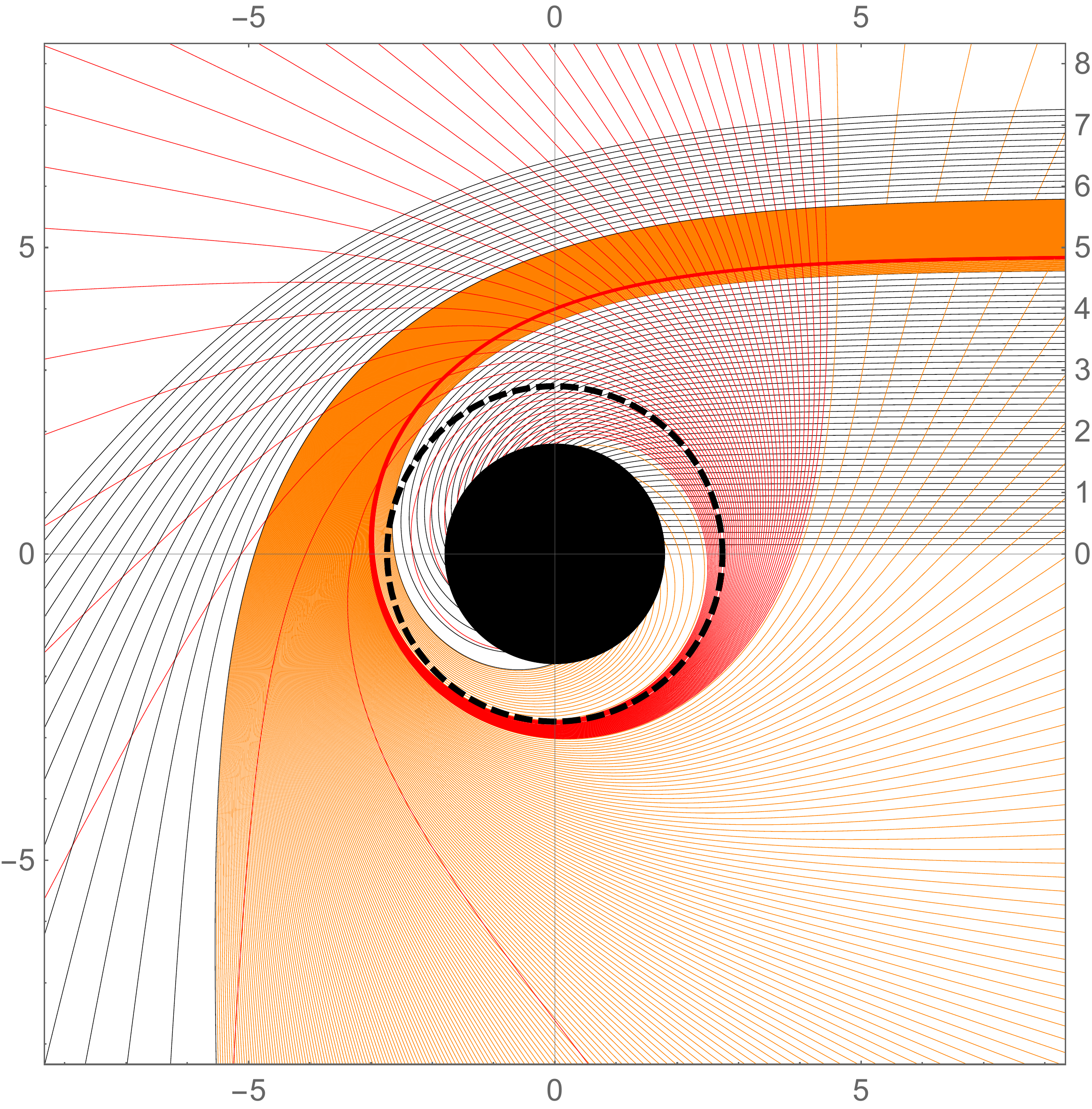}
	\includegraphics[width=0.22\textwidth]{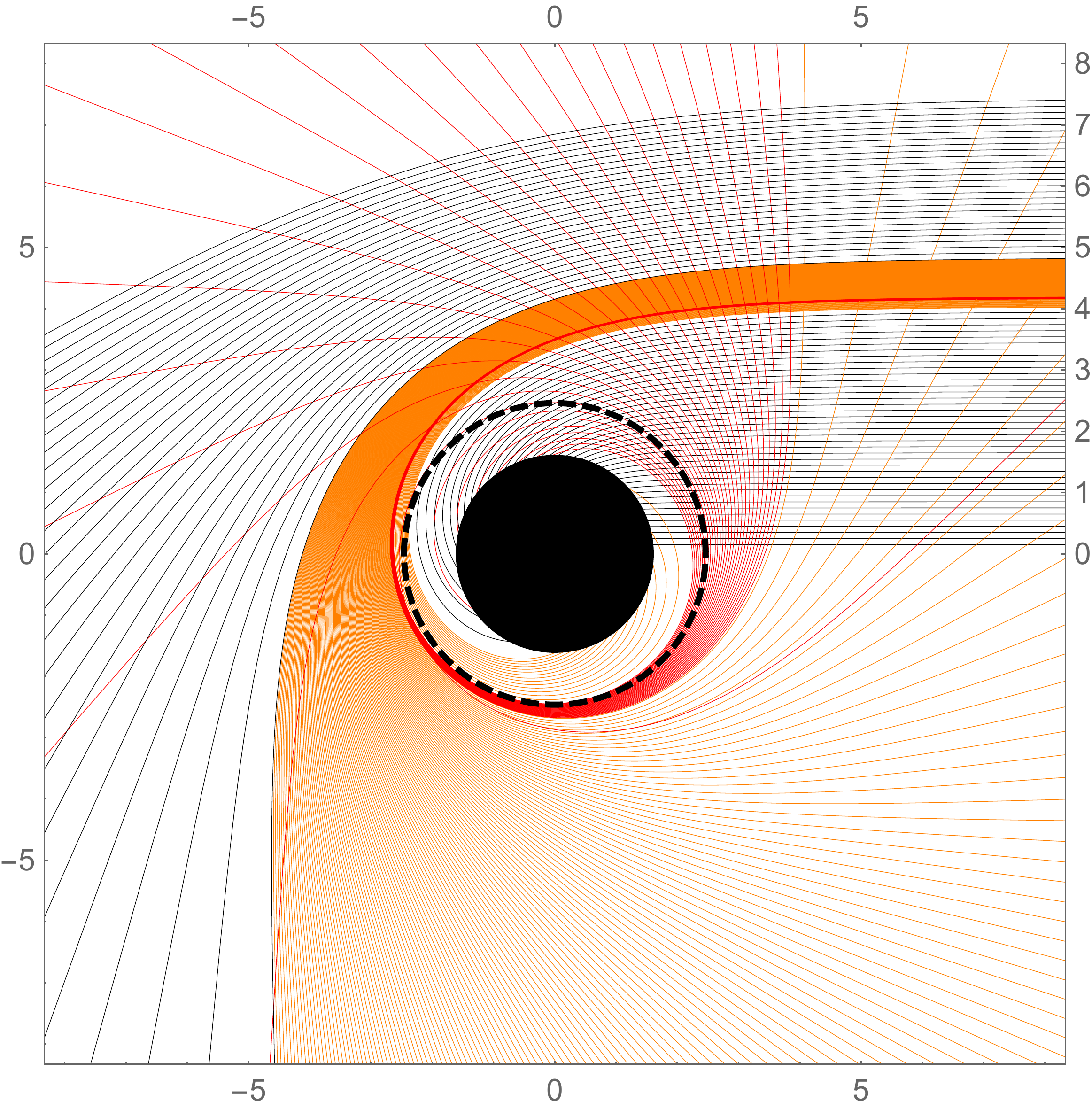}
	\includegraphics[width=0.22\textwidth]{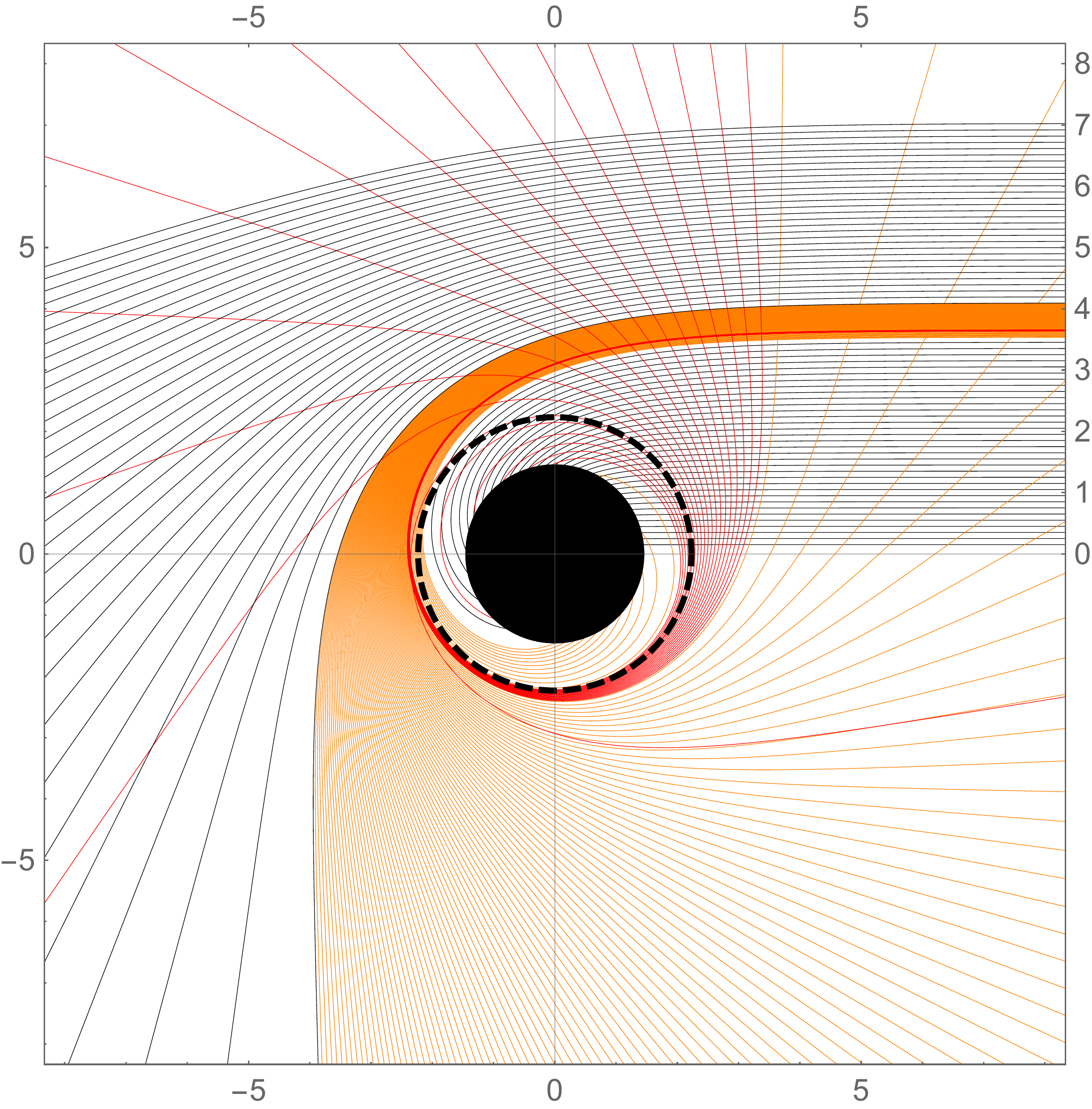}
	\includegraphics[width=0.22\textwidth]{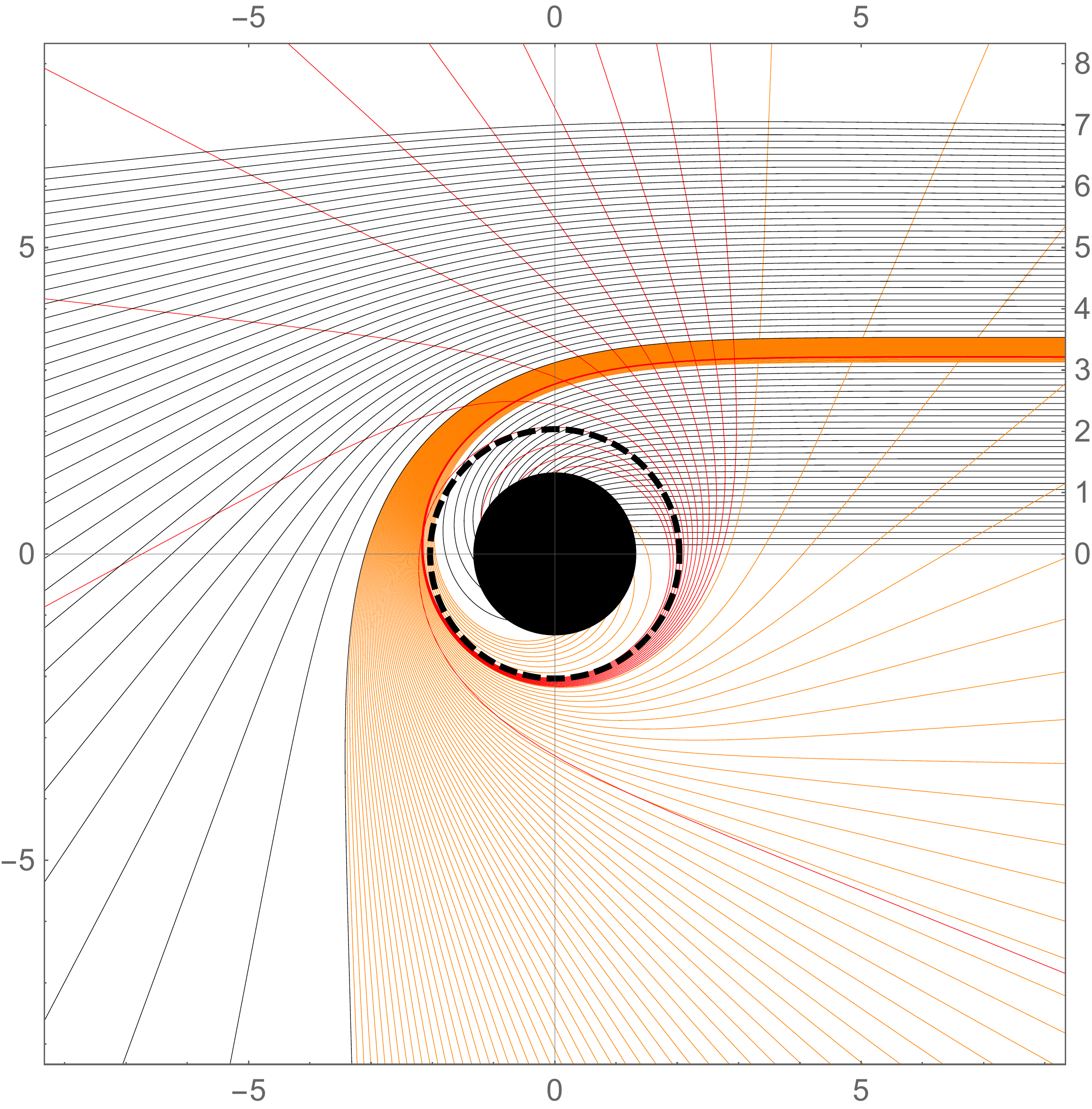}
	\caption{These are the null geodesic diagrams of photons under different values of the hairy parameter in polar coordinates. $Q_{YM}$ takes four values: 0 (top - left), 0.1 (top - right), 0.2 (bottom - left), and 0.3 (bottom - right). The black disc represents the black hole region within the event horizon, and the black dashed line represents the position of the unstable photon sphere radius. The observer is at the North Pole at an infinite distance from the black hole, and the various curved lines are the geodesic trajectories of photons. We use different colors to represent different types of geodesics: black represents the first type of geodesics that intersect the accretion disk only once; orange represents the second type of geodesics that intersect the accretion disk twice; red represents geodesics that intersect the accretion disk more than three times.}
	\label{6}
\end{figure}

It can be seen from Fig. \ref{6} that:
\begin{itemize}
	\item  The significance of the impact parameter in the figure is that it is the coordinate value corresponding to the null geodesic at infinity when it is directed to the right (north pole), that is, the radial distance from the center of the black hole of the light ray received by the observer at infinity.
	\item Under different hairy parameter conditions, the Einstein Maxwell power-Yang-Mills black hole and the RN black hole exhibit similar characteristics. The impact parameter intervals of the photon ring (represented by red curves) and the lensing ring (represented by orange curves) of both are relatively narrow, which can also be clearly seen from Table \ref{tab22}. In comparison, the impact parameter interval of the direct ring (represented by black curves) is relatively large and is composed of two mutually independent parts.
	\item As the Yang-Mills hairy parameter $Q_{YM}$ increases, it will cause the impact parameter intervals of the lensing ring and the photon ring in the spacetime to decrease. Correspondingly, the relevant brightness in these regions will also dim a bit.
	\item Under different hairy parameter situations, such a phenomenon will occur: As the impact parameter increases, the bending degree of the corresponding null geodesic gradually changes from small to large, reaches a maximum value at the position of the unstable photon sphere, and then gradually decreases. During this process, its color change follows the rule of black $\rightarrow$  orange $\rightarrow$  red $\rightarrow$  orange $\rightarrow$  black. 
	\item With the increase of the hairy parameter $Q_{YM}$, the event horizon, photon sphere radius, and critical impact parameter of the corresponding black hole will all change (gradually decrease), and the shadow area of the black hole will also decrease.
\end{itemize}

Therefore, the influence of the hairy parameter on photon trajectories cannot be ignored, and this characteristic can serve as a way to distinguish different black hole solutions.

After further calculations, the intervals corresponding to the impact parameters with different numbers of intersections with the accretion disk are determined, and the results are listed in Table \ref{tab22}, which also shows the results of the black hole event horizon $r_{h}$, photon sphere radius $r_{ph}$, critical impact parameter $b_{ph}$, and the innermost stable circular orbit $r_{isco}$(it will be used later). The innermost stable circular orbit $r_{isco}$can be calculated by Eq. \eqref{26}\cite{Gomez:2023qyv}.
\begin{equation}\label{26}
	\begin{split}
		r_{\text{isco}} = \frac{4M^4 - 3M^2Q^2(1 + Q_{YM}) + M^{\frac{4}{3}}u^{\frac{1}{3}}(2M^{\frac{4}{3}}+u^{\frac{1}{3}})}{M^{\frac{5}{3}}(1 + Q_{YM})u^{\frac{1}{3}}}
	\end{split}.
\end{equation}

\begin{table*}[ht]
	\centering

	\renewcommand{\arraystretch}{1.5} 
	\begin{tabular}{p{1.0cm}p{1.7cm}p{1.7cm}p{1.8cm}p{1.7cm}p{1.8cm}p{1.7cm}p{1.8cm}p{1.7cm}p{1.7cm}}
		\hline\hline 
		$Q_{YM}$ &  ${b}_{1}^{-}$ &  ${b}_{2}^{-}$ & ${b}_{2}^{+}$ & ${b}_{3}^{-}$ &${b}_{3}^{+}$ & $r_{h}$ & $r_{ph}$ &  $b_{ph}$ &$r_{isco}$\\
		\hline
		0   & 2.5792 & 4.6499 & 5.8482 & 4.8474 & 4.8957 & 1.8000 & 2.7369 & 4.8586 &5.4198 \\
		0.1 & 2.2969 & 4.0331 & 4.1723 & 4.2015 & 4.8319 & 1.6156 & 2.4613 & 4.1792 &4.8695 \\
		0.2 & 2.0620 & 3.5515 &3.6345  &3.6532  &4.0882  & 1.4613 &2.2310  &3.6390  &4.4099 \\
		0.3 & 1.8634 &3.1157  &3.5187  &3.1973  &3.2106  &1.3302  &2.0356  &3.2011  &4.0200 \\
		\hline\hline 
	\end{tabular}
	\caption{These are the values of the impact parameter $b$, the event horizon radius $r_{h}$, the photon sphere radius $r_{ph}$, the critical impact parameter radius $b_{ph}$, and the minimum orbit radius $r_{isco}$ when the hairy parameter $Q_{YM}$ takes different values, where the hairy parameter takes $Q_{YM}=0,0.1,0.2,0.3$ respectively.}
		\label{tab22}
\end{table*}
	
\subsection{The optical appearance of an Einstein Maxwell power-Yang-Mills black hole surrounded by a thin accretion disk}
This section will focus on the simulation research of optical characteristics, the total light intensity received by the observer, and the images of the photon ring and black hole shadow observed by the observer at infinity when the hairy parameter takes different values. As mentioned earlier, if the accretion disk is regarded as the only light source, for the null geodesics that intersect with the accretion disk, the light emitted from the accretion disk will propagate along these null geodesics to the position of the observer. It can be considered that energy is extracted each time the null geodesic intersects with the thin accretion disk. The more intersections there are, the more energy is extracted, and correspondingly, the brightness will increase, resulting in different contributions to the total intensity of the observed light, which directly affects the distribution of the observed intensity.

Considering that the thin accretion disk emits isotropically within the static frame, the specific intensity of the emission frequency $ \nu_{e} $ received by the observer is\cite{Wang:2023vcv}
\begin{equation}\label{27}
	\begin{split}
	I_0 (r, \nu_0) = g^3 I_e (r, \nu_e)
	\end{split}.
\end{equation}
Where \( g =\frac{\nu_0}{\nu_e} = \sqrt{f(r)} \) is the redshift factor, with \( \nu_0 \) and \( \nu_e \) representing the frequencies of the observed and emitted light, and \( I_0(r, \nu_0) \) and \( I_e(r, \nu_e) \) the specific intensities of the observed and emitted light, respectively. The total observed intensity over the entire wavelength band can be obtained by integrating $I_0 (r, v_0)$ over different frequency bands
\begin{equation}\label{28}
	\begin{split}
	I_{obs}(r) = \int I_0 (r, \nu_0) d\nu_0 = \int g^4 I_e (r, \nu_e) d\nu_e = f(r)^2 I_{em}(r)
	\end{split}.
\end{equation}
Among them, $I_{em}(r) = \int I_e (r, \nu_e) d\nu_e$ is the total emission intensity. Since the light extracts energy from the accretion disk each time it intersects with the accretion disk, as we mentioned before, light rays with the number of orbital turns $n > \frac{1}{4}$ will intersect with the accretion disk on the front side. When $n > \frac{3}{4}$, the light rays will bend around the black hole and intersect with the accretion disk for the second time on the back side. Further, when $n > \frac{5}{4}$, the light rays will intersect with the accretion disk for the third time on the front side again, and the total intensity we observe should be the sum of the energy extracted from each intersection\cite{Peng:2020wun,Gralla:2019xty}.
\begin{equation}\label{29}
	\begin{split}
	I_{obs}(b) = \sum_m f(r)^2 I_{em}(r) \Big|_{r = r_m(b)}
	\end{split}.
\end{equation}
Among them, we introduce the transfer function $r_m(b)$, changing its observed intensity to a function of $b$. It represents the radial coordinate of the  $m$-th intersection of the null geodesic with impact parameter $b$ and the accretion disk. The relationship between the number of intersections and the total angle bypassed by the light ray is $\varphi = \frac{2m - 1}{2} \pi$(According to the backward ray tracing method, the so-called $m$-th intersection is named in the order of successive intersections of the null geodesic from the observer with the accretion disk). In order to simplify the calculation, the absorption and reflection of light rays by the accretion disk are ignored here. Then, when $m$ is a fixed value, its slope $\frac{dr_m}{db}$ describes the demagnification factor, and a larger $m$ corresponds to strong demagnification.

 For the photon ring, there is a large strong demagnification, and its contribution to the total observed light intensity can be ignored. The contribution of the photon ring to the total light intensity is only a few percent of that of the lensing ring, so the total contribution of light intensity mainly comes from the lensing ring\cite{Gralla:2019xty}. The first three transfer functions $r_m(b) (m = 1, 2, 3)$ can be expressed as
\begin{equation}\label{30}
	\begin{split}
		r_1(b) = \frac{1}{u\left(\frac{\pi}{2}, b\right)}, \quad b \in (b_1^-, +\infty)
	\end{split},
\end{equation}
\begin{equation}\label{31}
	\begin{split}
	r_2(b) = \frac{1}{u\left(\frac{3\pi}{2}, b\right)}, \quad b \in (b_2^-, b_2^+)
	\end{split},
\end{equation}
\begin{equation}\label{32}
	\begin{split}
	r_3(b) = \frac{1}{u\left(\frac{5\pi}{2}, b\right)}, \quad b \in (b_3^-, b_3^+)
	\end{split}.
\end{equation}
Here, $u(\varphi, b)$ is the solution to the photon geodesic Eq. \eqref{21}. We have plotted the graphs of the first three transfer functions for different values of the Yang - Mills hairy parameter in Fig. \ref{7}.
\begin{figure}[h]
	\centering
	\includegraphics[width=0.22\textwidth]{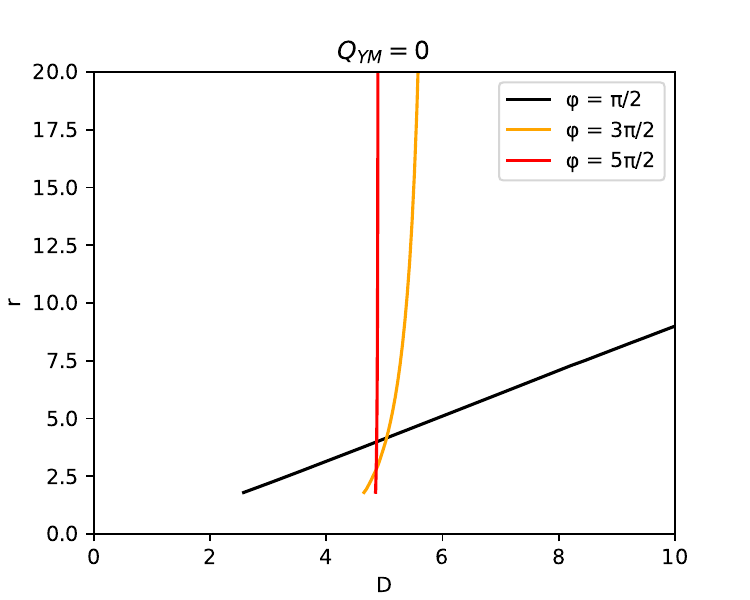}
	\includegraphics[width=0.22\textwidth]{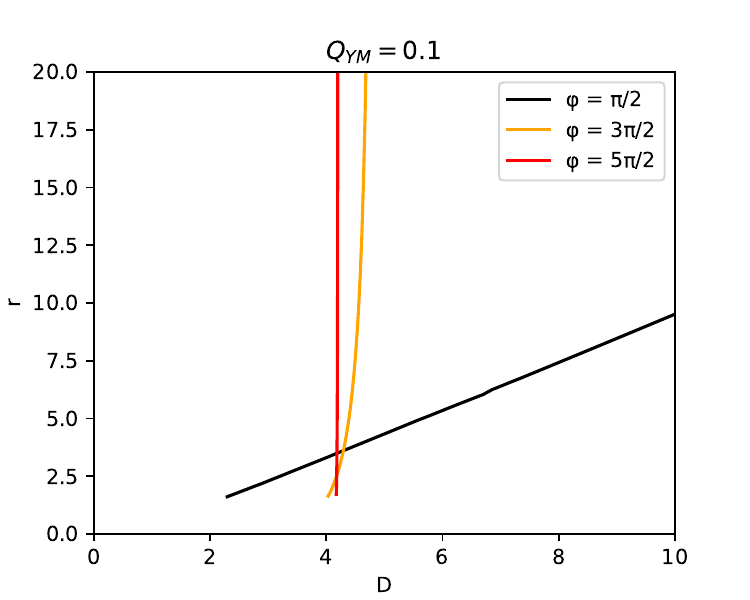}
	\includegraphics[width=0.22\textwidth]{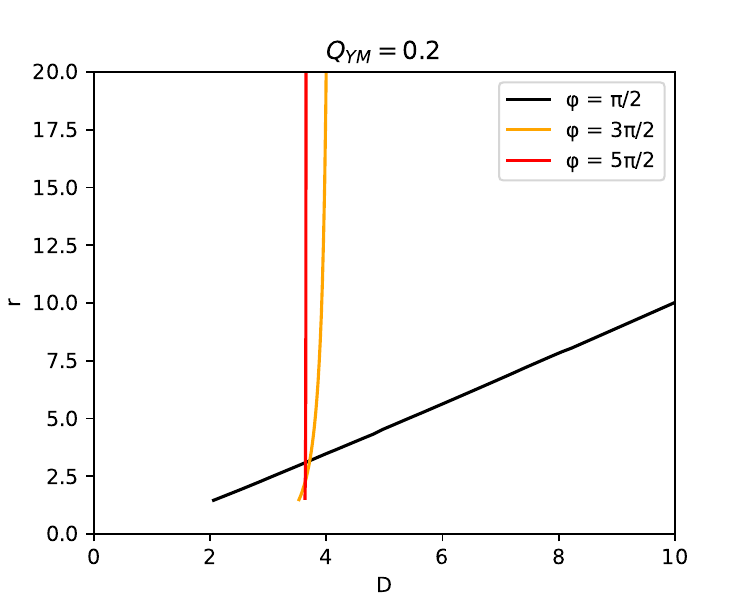}
	\includegraphics[width=0.22\textwidth]{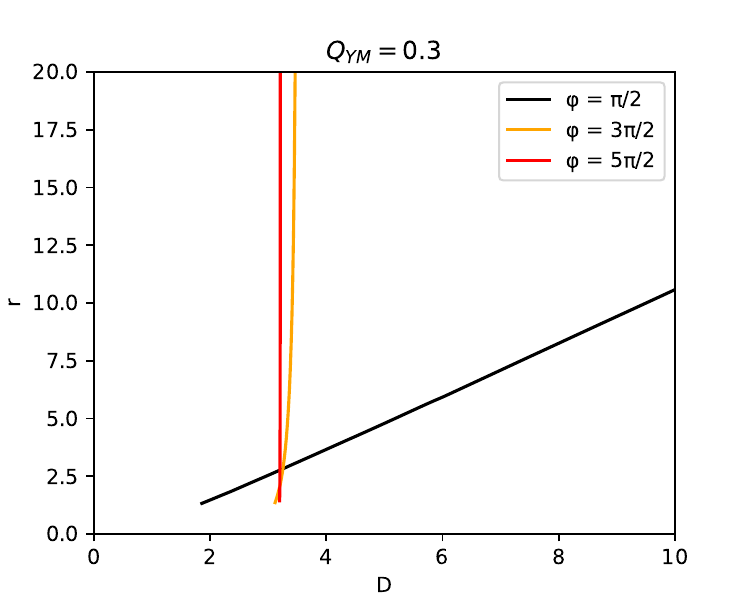}
	\caption{The images of the first three transfer functions for different values of the Yang-Mills hairy parameter, where the hairy parameters are taken as $Q_{YM}=0$,$Q_{YM}=0.1$,$Q_{YM}=0.2$,$Q_{YM}=0.3$ respectively. The curves of different colors represent the radial coordinates of intersections with the accretion disk once (black), twice (orange), and three times (red).}
	\label{7}
\end{figure}

From the figure, we can draw the following properties:
\begin{itemize}
	\item For the first transfer function (black curve) which can come from direct emission, lensing rings, and photon rings, we find that as the Yang-Mills hairy parameter increases, the initial value of the impact parameter $b$ decreases. Meanwhile, the slope of the black curve is close to 1, which means that the radial coordinate of the first intersection of the null geodesic with the accretion disk changes almost linearly with the impact parameter.
	\item For the second transfer function (orange curve) which can be derived from lensing rings and photon rings, we find that as the Yang-Mills hairy parameter increases, the value of the impact parameter $b$ also decreases. The increase in the slope indicates that its contribution to the light intensity is relatively small. 
	\item The third transfer function (red curve), which can only originate from the photon ring, has been studied, and it was found that as the Yang-Mills hair parameter increases, the threshold of its collision parameter decreases, and the transfer function exhibits a larger slope. However, in terms of light intensity contribution, it is smaller than the second transfer function and can almost be ignored.
\end{itemize}

In order to further verify the prediction of the transfer function, we must conduct an in-depth study on the contributions of direct emission, lensing rings, and photon rings to the total intensity of the light and the optical appearance image of the observed Einstein Maxwell power-Yang-Mills black hole under specific emission intensity conditions. For this purpose, we have considered three specific toy model emission functions.
For the first emission model, we assume that the radiation intensity of the accretion disk is a function that decays in a quadratic power form starting from the innermost stable orbit.
\begin{equation}\label{33}
	\begin{split}
		 I_{em1}(r)=\begin{cases}
			I_{0}\left[\frac{1}{r-(r_{isco}-1)}\right]^{2},&r > r_{isco}\\
			0,&r\leq r_{isco},
		\end{cases}
	\end{split},
\end{equation}
Here, $I_{0}$ represents the maximum emission intensity (the same applies hereinafter). The data of the radius $r_{isco}$ of the innermost stable orbit is given in Table \ref{tab22}. For the second emission model, we assume that the radiation intensity of the accretion disk is a function that decays in a cubic power form starting from the position of the photon sphere orbit.
\begin{equation}\label{34}
	\begin{split}	
		I_{em2}(r)=\begin{cases}
			I_{0}\left[\frac{1}{r-(r_{ph}-1)}\right]^{3}, & r > r_{ph}\\ 			
			0, & r \leq r_{ph}
		\end{cases}
	\end{split},
\end{equation}
Here,$r_{ph}$represents the photon sphere radius. For the third emission model, we take into account a more slowly decaying function which starts from the position of the event horizon radius $r_{h}$.
\begin{equation}\label{35}
	\begin{split}	
	I_{em3}(r)=\begin{cases}
		I_{0}\frac{\frac{\pi}{2}-\arctan\theta\left[r-(r_{isco}-1)\right]}{\frac{\pi}{2}-\arctan\theta\left[r_{h}-(r_{isco}-1)\right]}, & r > r_{h}\\
		0, & r \leq r_{h}
	\end{cases}
	\end{split}.
\end{equation}

Considering that in the real universe, there are always various substances surrounding black holes, it is necessary to further study the observational images under different circumstances. Therefore, next we will substitute these three different emission intensity functions and transfer functions into Eq. \eqref{27} to study the total emission and the total observed intensity under the three toy models when the Yang - Mills hairy parameter takes the values of 0, 0.1, and 0.2, as well as their optical appearances (see Figs. \ref{8}, \ref{9}, and \ref{10}). The first column of each figure shows the total emission intensity, the second column displays the total observed intensity, that is, the contributions of the three ring structures to the total light intensity. The third column presents the variation of the total observed intensity with the impact parameter and converts it into the corresponding two - dimensional image, thus generating the optical appearance of the accretion disk.

\begin{figure*}
	\includegraphics[width=1\textwidth]{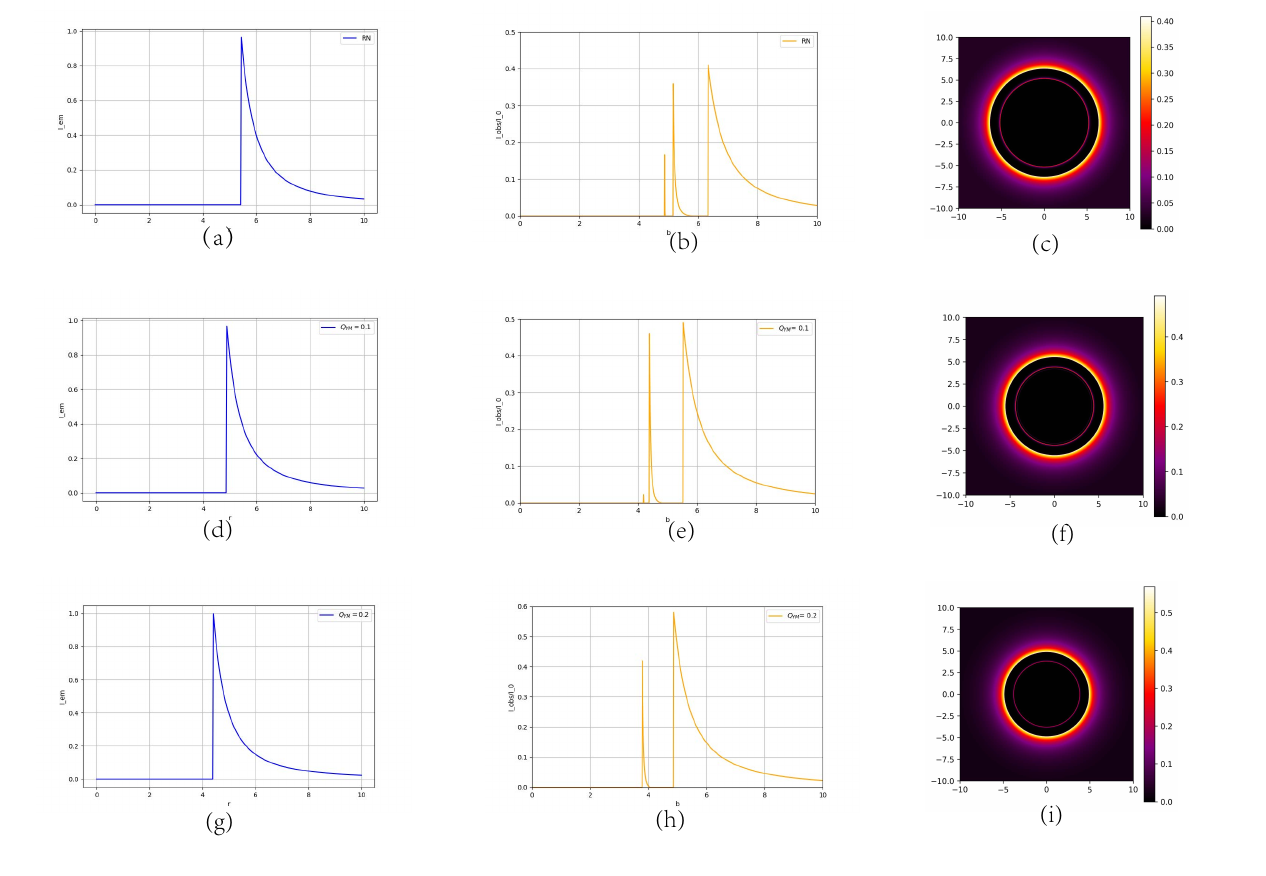}
	\caption{Images of the first toy model when the Yang-Mills hairy parameter $Q_{YM}$ takes the values of 0, 0.1, and 0.2 respectively. The first column shows the emission intensity $I_{em1}(r)$, the second column presents the total intensity $I_{obs}(b)$ observed by the observer, and the third column displays the optical appearance seen by the observer.}
	\label{8}
\end{figure*}

As shown in Fig. \ref{8}, it is the emission and reception intensities and the optical appearance of the first toy model:  
\begin{itemize}
	\item In the first toy model, the emission intensity in the first column exhibits a specific pattern: it reaches its maximum value at the radius $r_{isco}$ of the innermost stable circular orbit and then decays rapidly. Moreover, as the hairy parameter $Q_{YM}$ continuously increases, the peak of the emission intensity moves in the direction of decreasing $b$ value. This implies that when the Yang - Mills hairy parameter increases, the minimum radius of photons around the black hole decreases accordingly. 
	\item Regarding the total intensity observed by the observer in the second column, as shown in Figure (b), the three peaks from left to right are the photon ring, the lensing ring, and the direct ring. However, the region of the photon ring is extremely narrow, which can be seen in Figure (b). As the hairy parameter  $Q_{YM}$ increases, the peak of the ring moves in the direction of decrease. Therefore, in the optical appearance diagrams with larger values of  $Q_{YM}$, the radius of the ring we observe is smaller. When $Q_{YM}=0.2$, the peak of the photon ring is almost invisible, so its contribution to the total observed intensity is almost negligible. The fact that the total observed intensity has two distinct peaks indicates that when we project the total observed intensity onto a two dimensional plane, the corresponding optical appearance should be two bright rings, which can also be clearly seen in the optical appearance in the third column.  
	\item In the dark area of Figure (c) in the third column, there is a very thin bright ring, which is the lensing ring. There is an even thinner photon ring on the inner side of the lensing ring. Its brightness is too low to be easily observed, and the contributions of these two rings to the overall brightness are extremely limited. The direct ring is located outside the lensing ring. At the position of the direct ring, the brightness suddenly increases and becomes very bright. Subsequently, the brightness gradually decreases as it moves further outwards.  
	\item  In conclusion, under different hairy parameters  $Q_{YM}$, the positions of the peak values of the total intensity observed by the observer will change. As the Yang-Mills hairy parameter keeps increasing, the peak of the observed total intensity moves in the direction of decreasing value $b$, which causes the radius of the optical appearance to change accordingly. In the Einstein Maxwell power-Yang-Mills black hole, precisely because of this characteristic, there is no degeneracy phenomenon in its optical appearance images. It is worth noting that the values of the Yang-Mills hair corresponding to the photon rings with different radius in the optical appearance of this black hole are different. This rule provides an effective means to test the Yang-Mills hair, which is conducive to our further exploration of the physical phenomena and essence related to this black hole.  
\end{itemize}
\begin{figure*}
	\includegraphics[width=1\textwidth]{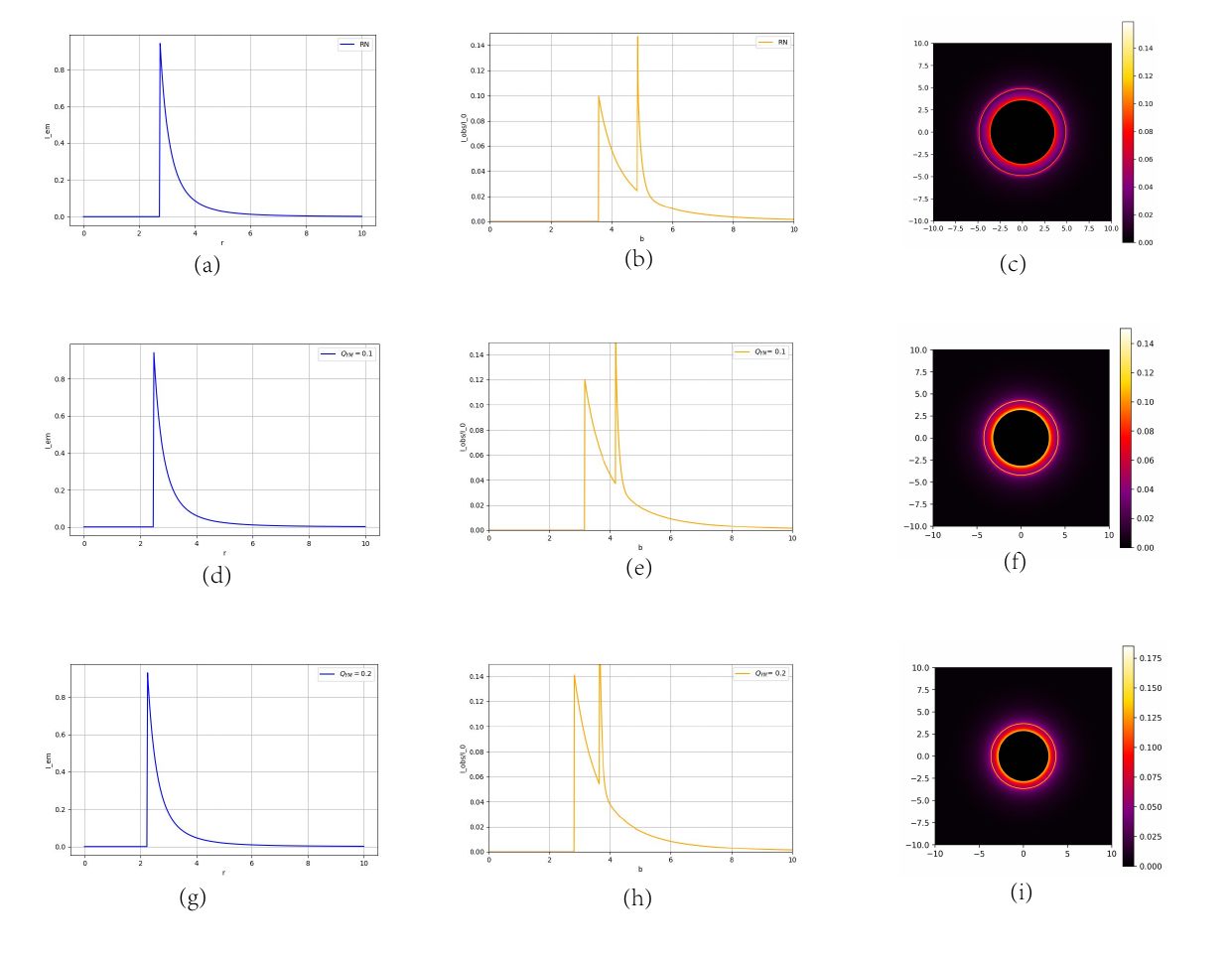}
	\caption{Images of the second toy model when the Yang-Mills hairy parameter $Q_{YM}$ takes the values of 0, 0.1, and 0.2 respectively. The first column shows the emission intensity $I_{em2}(r)$, the second column presents the total intensity $I_{obs}(b)$ observed by the observer, and the third column displays the optical appearance seen by the observer.}
	\label{9}
\end{figure*}

The emission and reception intensities and the optical appearance of the second toy model are shown in Fig. \ref{9} :

\begin{itemize}
	\item In the first column, the emission intensity of the second toy model reaches its maximum value at the photon sphere $r_{ph}$, and then decays rapidly. Moreover, as the Yang-Mills hairy parameter $Q_{YM}$ increases, the peak moves in the direction of decrease $b$. That is, an increase in the Yang - Mills hairy parameter $Q_{YM}$ can reduce the radius of the photon sphere of photons around the black hole.  
	\item For the total intensity observed by the observer in the second column, as in (b), the two peaks from left to right are respectively a peak formed by the direct ring and another peak formed by the superposition of the photon ring and the lensing ring. Different from the first model, the direct ring appears inside the lensing ring and the photon ring. As  $Q_{YM}$ increases, the peaks of the rings also move in the direction of decreasing $b$, so the radius of the rings seen in the optical appearance diagrams with larger values of  $Q_{YM}$ will also be smaller.  
	\item Judging from the total observed intensity presented in the second column, the total observed intensity is decreasing compared to the first toy model. This can also be observed from the optical appearance in the third column. Precisely because of this, the optical appearance also looks relatively darker. In addition, the brighter direct ring is located inside the photon ring and the lensing ring.  
	\end{itemize}
\begin{figure*}
	\includegraphics[width=1\textwidth]{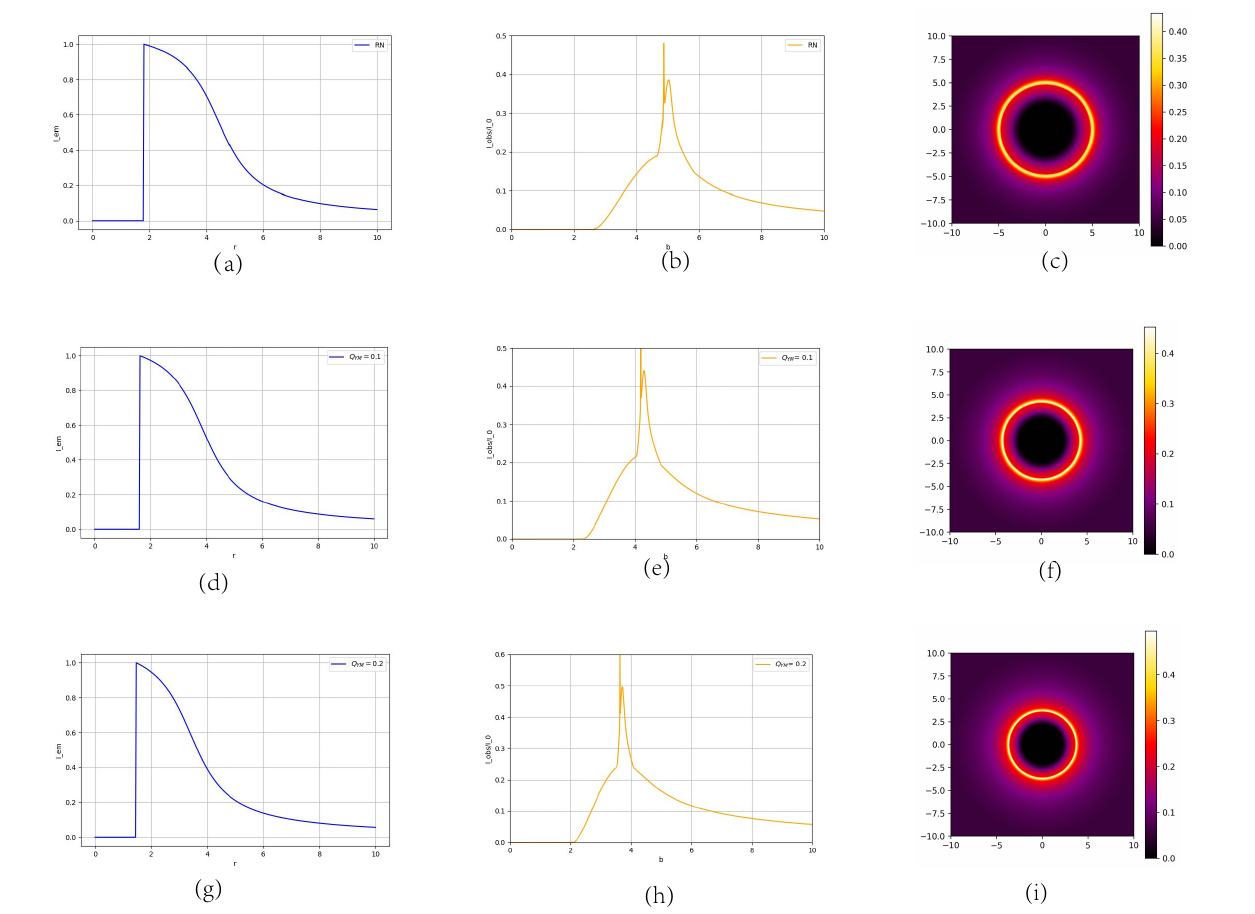}
	\caption{Images of the second toy model when the Yang-Mills hairy parameter $Q_{YM}$ takes the values of 0, 0.1, and 0.2 respectively. The first column shows the emission intensity $I_{em3}(r)$, the second column presents the total intensity $I_{obs}(b)$ observed by the observer, and the third column displays the optical appearance seen by the observer.}
	\label{10}
\end{figure*}

Fig. \ref{10} shows the emission and reception intensities as well as the optical appearance of the third toy model:
\begin{itemize}
	\item The emission intensity of the third toy model in the first column reaches its maximum value at the radius of the event horizon $r_{h}$ and then decays slowly. Moreover, as the Yang-Mills hairy parameter $Q_{YM}$ increases, the peak value $b$ shifts in the direction of decreasing. That is to say, the same result also holds for the third toy model—an increase in the hairy parameter $Q_{YM}$ can reduce the photon sphere radius of photons around the black hole.  
	\item From the graph (b) of the total observed intensity for the observer in the second column, it can be seen that there is an obvious peak, which appears at the position of the photon ring, indicating that the intensity of the photon ring plays a decisive role in the maximum value of the total observed intensity. As $Q_{YM}$ increases, a second peak appears at the position of the lensing ring. The difference in the intervals between these two peaks shows that the lensing ring contributes more to the total flux than the photon ring. Compared with the second toy model, the observed intensity is slightly enhanced. As $Q_{YM}$ increases, the peak gradually becomes stronger, and the position of the peak shifts in the direction of decreasing $b$. Therefore, the radius of the ring seen in the optical appearance diagram for larger values of $Q_{YM}$ will also be smaller. 
	\item Compared with the first and second models, the third model has only one ring that changes from dark to bright and then from bright to dark, while the first and second models have two rings. Moreover, due to the greater total observed intensity in the second column, the third model has a higher brightness. In fact, the third model should also have two rings. However, judging from the observed intensity and the image, the distance between the two peaks is too small, so it is very difficult to discern the double-ring structure.  
\end{itemize}

In conclusion, the hairy parameter $Q_{YM}$ has an impact on the observed intensity under different models. The larger the hairy parameter, the stronger the observed intensity. Different hairy parameters also affect the position of the peak of the observed intensity, thus influencing the optical appearance image. Therefore, these results can serve as effective characteristics for distinguishing black holes with different hairy parameters. 
\section{Conclusion}\label{diwujie}
In this paper, the null geodesics near the Einstein Maxwell power-Yang-Mills black hole are studied. Meanwhile, based on the distribution of light rays, the optical images of the black hole surrounded by an optically and geometrically thin accretion disk are calculated, and the influence of the hairy parameter on both is discussed.  

Firstly, the Einstein Maxwell power-Yang-Mills black hole solution is briefly reviewed. In this solution, a p-power Yang-Mills term and a hairy parameter related to non-Abelian charge are introduced, and the influence of the hairy parameter on the radius of the event horizon is analyzed. By analyzing the change of the hairy parameter $Q_{YM}$  when $Q=0.6$, it is found that a critical situation occurs for the black hole when $Q_{YM}=1.77778$; when $Q_{YM}<1.77778$, the Einstein Maxwell power-Yang-Mills black hole has two horizons, corresponding to the Cauchy horizon and the event horizon respectively; when $Q_{YM}>1.77778$, there is no event horizon and it is a naked singularity situation. Due to the cosmic censorship hypothesis proposed by Penrose, which prohibits the appearance of naked singularities, in this paper, we mainly consider the case where $Q_{YM}<1.77778$.

Secondly, the equations of motion trajectories and the effective potential of photons around this black hole were studied. Starting from the Lagrangian of the black hole, the geodesic equations and the effective potential of photons were obtained. Based on the effective potential, the radius of the unstable photon sphere orbit could be determined. Subsequently, the null geodesics around the hairy black hole were explored through the backward ray-tracing method. It was found that the hairy parameter $Q_{YM}$ had an impact on both the distribution and classification of geodesics. Moreover, when the hairy parameter increased, the event horizon radius $r_{h}$, photon sphere radius $r_{ph}$,  the radius of the innermost stable circular orbit  $r_{isco}$ and critical impact parameter $b_{ph}$ of the black hole would all decrease. The interval of the impact parameters corresponding to the photon ring and the lensing ring would also become narrower, which would further affect the image of the null geodesics of the black hole. The data and images could be seen in Table \ref{tab22} and Fig. \ref{6} respectively. 

Finally, the total emission intensity, the total observed intensity, and the optical appearance corresponding to each of the three simplified models with different values of the hairy parameter $Q_{YM}$ under the irradiation of optically and geometrically thin accretion disks were studied. It was found that as the hairy parameter increased, the total observed intensity of each model would become stronger, and all the peaks of the total observed intensity were shifting in the direction of decreasing $b$, leading to a reduction in the ring radius. In addition, all three emission models showed that the direct image dominated the total flux, the lensing ring was a bright yet extremely narrow ring that was difficult to observe, and the contribution of the photon ring was negligible and almost invisible. In conclusion, different hairy parameters would result in different optical appearances, which also differed from those of the standard RN black hole. Further analysis led to the conclusion that when the hairy parameters were different, there was no degeneracy in their optical appearances and geodesics. Therefore, it could be considered that the geodesics and optical appearance images of photons could be used to distinguish different spacetime properties. Theoretically speaking, it was possible to distinguish the spacetime metrics of Einstein Maxwell power-Yang-Mills black holes with different hairy parameters. This characteristic creates a potential opportunity for characterizing various black hole solutions by observing the photon rings of black holes in the future, thus providing a valuable research approach and direction for delving deeper into the mysteries of black holes.

\section*{Acknowledgements}
We acknowledge the anonymous referee for a constructive report that has significantly improved this paper. This work was supported by Guizhou Provincial Basic Research Program (Natural Science) (Grant No. QianKeHeJiChu-[2024]Young166), the Special Natural Science Fund of Guizhou University (Grant No.X2022133), the National Natural Science Foundation of China (Grant No. 12365008) and the Guizhou Provincial Basic Research Program (Natural Science) (Grant No.QianKeHeJiChu-ZK[2024]YiBan027 and QianKeHeJiChu-ZK[2025]General Program680.).

\bibliography{ref}
\bibliographystyle{apsrev4-1}

\end{document}